\documentclass[
aps,nofootinbib,
showpacs,showkeys,preprint
tightenlines,preprintnumbers,] {revtex4}

\usepackage{epsf,epsfig,subfigure,graphicx,amsmath,amssymb 
}
\usepackage{color}
\newcommand{\dis}[1]{\begin{equation}\begin{split}#1\end{split}\end{equation}}
\newcommand{\ar}[3]{#1^{+#2}_{-#3}}
\newcommand{\ie}{{\it i.e.~}}
\newcommand{\etal}{{\it et al.\,}}

\newcommand{\Qem}{Q_{\rm em}}

\newcommand{\Z}{{\bf Z}}

\newcommand{\delks}{\alpha_{_{\rm KS}}}
\newcommand{\delpdg}{\alpha_{_{\rm PDG}}}
\newcommand{\delksl}{\alpha_{_{\rm KS}}^{\ell}}

\newcommand{\dell}{\delta_{_{\rm PMNS}}}

\def\sw0{{$\sin^2\theta_W^0$}}

\def\CP{${\cal CP}$}

\def\E6{{\rm E_6}}

\def\EE8{{\rm E_8\times E_8'}}

\begin{document}
 
\draft

\title{\Large\bf Jarlskog determinant and data on  flavor matrices}

\author{Jihn E.  Kim, Se-Jin Kim, Soonkeon Nam,  Myungbo Shim}

\address{
Department of Physics, Kyung Hee University, 26 Gyungheedaero, Dongdaemun-Gu, Seoul 02447, Republic of Korea 
}

\begin{abstract} 
The essences of the weak \CP\,violation, the quark and lepton Jarlskog invariants, are determined toward future model buildings beyond the Standard Model (SM). The equivalence of two calculations of  Jarlskog invariants gives a bound on the \CP\,phase in some parametrization.
Satisfying the unitarity condition,  we obtain the CKM and MNS matrices from the experimental data, and present the results in matrix forms.  The  Jarlskog  determinant  $J^q$ in the quark sector is found to be $\sim 3.11\times 10^{-5}|\sin\delks|$  while $J^\ell$ in the leptonic sector is  $\sim 2.96\times 10^{-2}|\sin\delksl|$ in the normal hierarchy parametrization.
 
\end{abstract}
\maketitle
  
 \section{Introduction}

Ever since the discovery of  weak \CP\,violation \cite{Cronin64}, its origin and  cosmological implication have  been a mystery. Ideas such as a tiny \CP\,violation effect in the strong interaction sector or scalar mediated weak \CP\,violation had not been considered any more as  leading ones after  Kobayashi and Maskawa(KM) found that three left-handed(L-handed) charged currents lead to weak \CP\,violation effects \cite{KM73}.  In early 1960's, \CP\,violation had been an interesting topic even to laymen \cite{Feynman65}.   In late 1960's, \CP\,violation had been considered as an indispensable ingredient in baryogenesis, creating baryons  out of a  baryonless universe \cite{Sakharov67}. 

\CP\,violation can arise in three varieties, (1) the strong \CP\,violation \cite{StrongCP}, (2) the weak \CP\,violation, and (3) \CP\,violation by singlets beyond the Standard Model (BSM). The strong \CP\,problem has led to the so-called axion physics which is one of the leading candidates for dark matter in the Universe \cite{KimRMP10} but short of explaining the current baryon asymmetry in the universe.  Out of the remaining two, the leading candidate toward the baryon asymmetry is the   \CP\,violation by the BSM singlets. On the other hand, the weak \CP\,violation hints a crucial information on a fundamental theory of elementary particles. The reason is the following. Because the strangeness changing neutral current effects are strongly suppressed \cite{GIM70}, the flavor changing effects are dominated by flavor changing charged currents. In the SM, the L-handed doublets encode this information. The observation of the weak \CP\,violation \cite{Cronin64} requires three or more L-handed SM doublets. If we supersymmetrize the SM and   require  asymptotic freedom above the TeV scale, four or more L-handed doublets are forbidden. Thus, three L-handed doublets are unique. This observation leads to the following flavor puzzles.

The flavor puzzle in the SM constitutes in two parts: (i) ``Why are there three chiral families?'', and (ii) ``Why is the Cabibbo--Kobayashi--Maskawa(CKM) matrix \cite{Cabibbo63,KM73} almost diagonal while it is not so in the Pontecorvo--Maki--Nakagawa--Sakata(PMNS) matrix \cite{PMNS1, PMNS2, PMNS3}?''  Here, we suggest the usefulness of Jarlskog determinant \cite{Jarlskog85} answering the second flavor problem if three families are given. Because string theory has been believed to be sufficiently restrictive  below the string scale, works on three families from string compactification   exploded under the phrases `standard-like models in string compactification' \cite{KimPLB19ga}  and `SUSY GUTs from string' \cite{AEHN87,KimKyae07,Huh09,SU7}. All these models attempted to realize three chiral families.  But, the more difficult problem is (ii) on the CKM matrix. Because of the reduction of the number of Yukawa couplings in GUTs compared to the standard-like models, an anti-SU(5) has been attempted for the flavor problem  \cite{KimPRD18fl}. For a successful phenomenology, not only the Yukawa couplings but also the discrete symmetry $\Z_{4R}$ \cite{Leeetal11} and a mechanism for SUSY breaking at an intermediate scale  \cite{KimPLB84} are needed. GUTs help analyzing these issues also  \cite{KimPRDz4R,KimPLB19ga}. 

It is known that using any CKM parametrization leads to the same physical quantities. In particular, the Jarlskog invariants  $J^q$ of the quark sector and $J^\ell$ of the lepton sector must be the same in any parametrization. In this paper, in a global fitting using the 5 allowable real-number data points in the CKM and PMNS matrices, we will find that the quark sector $J^q$ is of order $10^{-5}$  both in the Kim--Seo(KS) parametrization \cite{KimSeo11} and in the PDG book parametrization \cite{PDG18}.  
Thus, we draw the attention that the Jarlskog determinants $J^q$   and $J^\ell$ are useful in analyzing the actual data. In fact, using the allowable  $J^q$ and $J^\ell$, we obtain numerical $3\times 3$ CKM and PMNS matrices at our best capability, which can be easily applied in the future BSM model buildings. We apply our model-independent analysis to determine the elements of PMNS matrix. Consistency of $J^\ell$ in two parametrizations  \cite{KimSeo11,PDG18} allows us to express the PMNS matrix without the information on the PMNS phase $\dell$.  
We hope that our anayses on the   CKM and PMNS matrices can be useful in the future model building.

In Sec. \ref{sec:Definition}, we define parametrizations used in the paper. In Sec. \ref{sec:J}, the essence of the Jarlskog determinant is presented. In Sec. \ref{sec:Data}, we analyse the PDG data  presented in matrix forms with absolute values. For some specfific exclusive process may give some angle with a smaller error bar  than presented in the matrix form, but such anayses must assume some value on the phase. We try to use the PDG matrix without phase information because we want to present the final result with the undetermined phase.
Sec. \ref{sec:Conclusion} is a conclusion.

 \section{A useful parametrization}\label{sec:Definition}

Because the flavor changing neutral current   effects are almost absent, the structure of three  L-handed  doublets of quarks and leptons are enough for the flavor study in the SM,
 \begin{eqnarray}
&& \left(
\begin{array}{c}
u^{\prime\alpha}\\ d^\alpha
\end{array}
\right)_L,~  \left(
\begin{array}{c}
c^{\prime\alpha}\\ s^\alpha
\end{array}
\right)_L,~  \left(
\begin{array}{c}
t^{\prime\alpha}\\ b^\alpha
\end{array}
\right)_L \label{eq:SMfamiliesQ}
\\[0.5em] 
 && \left(
\begin{array}{c}
\nu^{\prime}_e\\ e
\end{array}
\right)_L,~~ \left(
\begin{array}{c}
\nu^{\prime}_\mu\\ \mu
\end{array}
\right)_L,~ ~ \left(
\begin{array}{c}
\nu^{\prime}_\tau\\ \tau
\end{array}
\right)_L. \label{eq:SMfamiliesL}
 \end{eqnarray} 
 In Eqs. (\ref{eq:SMfamiliesQ}) and (\ref{eq:SMfamiliesL}), we choose bases such that the lower components are the mass eigenstates while the upper components are mixtures of the mass eigenstates.
 
We use the Kim--Seo(KS) form  for the CKM matrix \cite{KimSeo11},
 \begin{equation}
V^{\rm KS}= \left(\begin{array}{ccc} c_1,&s_1c_3,& s_1s_3  \\ [0.2em]
-c_2s_1, & c_1c_2c_3+ s_2s_3 e^{-i\delks},&c_1c_2s_3-s_2c_3e^{-i\delks}  
 \\[0.2em]
 -s_1s_2e^{+i\delks},&  -c_2s_3 +c_1s_2c_3 e^{+i\delks}, & c_2c_3 +c_1s_2s_3 e^{+i\delks}
\end{array}\right),\label{eq:KSformM}
 \end{equation}
where  $c_i$ and $s_i$ are  cosines and sines of three real angles $\theta_i\,(i=1,2,3)$  and $\delks$ is a \CP\,phase. The KS form is written such that the elements in the 1st row are all real, which makes it easy to draw the Jarlskog triangle with one side sitting on the horizontal axis. 
 For the MNS matrix of Eq. (\ref{eq:SMfamiliesL}), we use another four parameter set,
$\Theta_i$ (giving corresponding cosine $C_i$ and sine $S_i$) and $\delksl$.

The situation with the other forms of parametrization is the following.
The original KM form  \cite{KM73} gives a complex determinant. The Maiani form \cite{Maiani76} is not exactly unitary, and the Wolfenstein form is designed to be approximate \cite{Wolfenstein83}. A relevant one, being unitary with a real determinant,  is the form used in the PDG book \cite{PDG18ckm} which is the form originally used by Chau and Keung \cite{CK85}. The PDG form has four real elements, (11), (12), (23), and (33) elements.   

In this paper, we determine $J$ explicitly in the KS and the PDG  forms, using 5 real numbers: (11),  (12),  (13),  (21), and $-|(31)|$ with the KS form, and  (11),  (12), $+|(13)|$,  (23),  and (33)  with the PDG form. For the CKM matrix, we determine the \CP\,phase of $\sim 90^{\rm o}$ in the KS form and  $\sim 70^{\rm o}$ in the PDG form.  A choice of parametrization can be   preferred depending on how easily such \CP\,phases result in some ultra-violet completed theories.  The leptonic \CP\,phase has large error bars at present.  
For the readers' convenience, we present the PDG form of the CKM matrix \cite{PDG18ckm} here. 
\begin{equation}
V^{\textrm{PDG}}=\left(
\begin{array}{ccc}
 c_{12} c_{13} & s_{12}c_{13} & s_{13}e^{-i \delpdg} \\
-s_{12}c_{23}-c_{12}s_{23}s_{13}e^{i\delpdg} & c_{12}c_{23}-s_{12}s_{23}s_{13}e^{i\delpdg} & s_{23}c_{13} \\
s_{12}s_{23}-c_{12}c_{23}s_{13}e^{i\delpdg} & -c_{12}s_{23}-s_{12}c_{23}s_{13}e^{i\delpdg} & c_{23}c_{13} \\
\end{array}
\right)      
\end{equation}

 \section{Usefulness of Jarlskog determinant}\label{sec:J}
\begin{figure}[!t]
\hskip 0.01cm \includegraphics[width=0.4\textwidth]{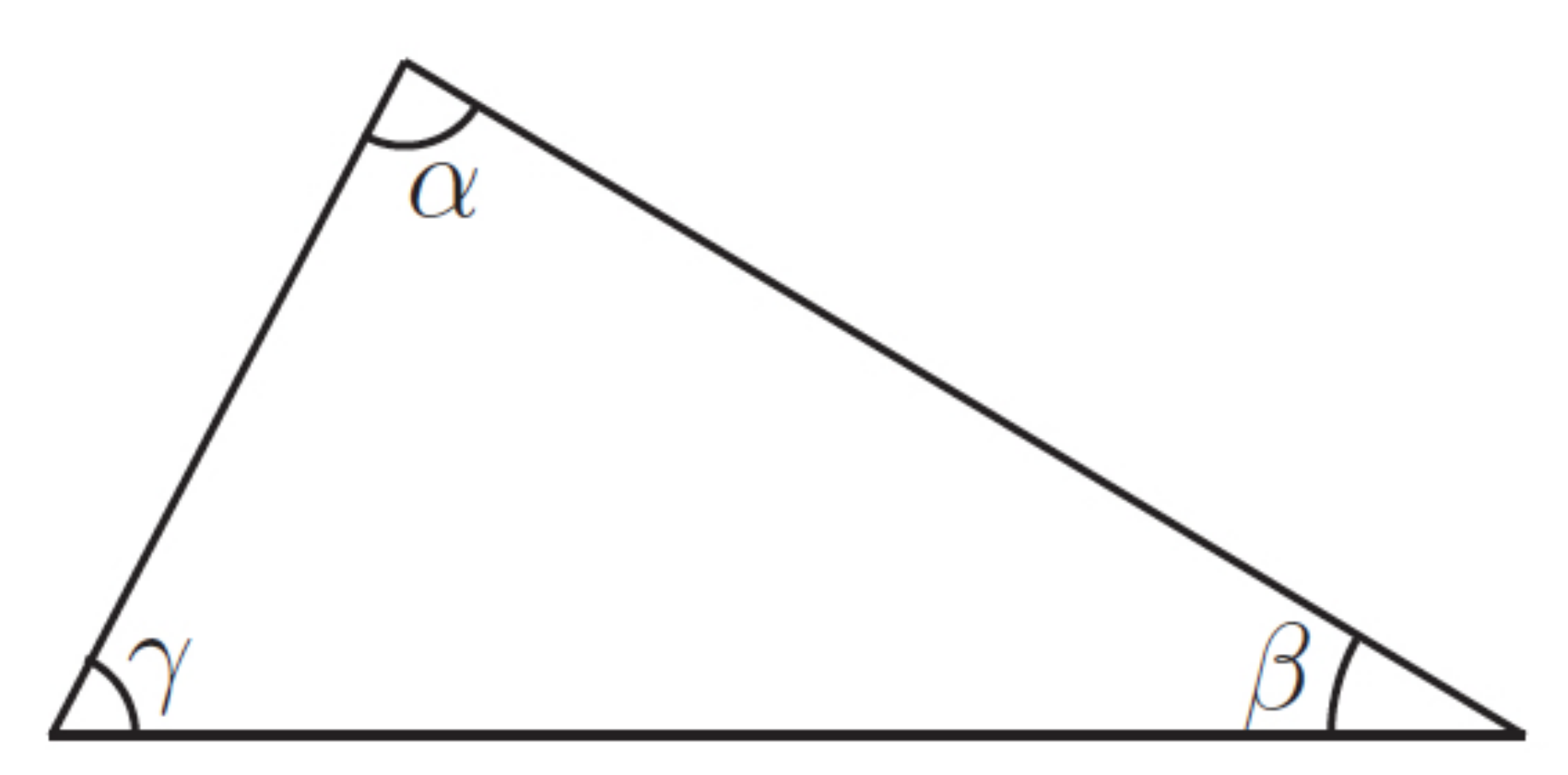}
\caption{The layout of the Jarlskog triangle in the PDG book \cite{PDG18ckm}. }\label{fig:JTriPDG}
\end{figure}

 The \CP\,violation effect\textcolor{red}{s} for transitions involving $c, t, d$, and $s$ were parametrized \cite{Jarlskog85}, here rearranging the original expression  of \cite{Jarlskog85}, as $2\,{\rm Im\,}(V_{22}V_{21}^*)^* (V_{32}V_{31}^*) / (|V_{21}V_{32}|^2+|V_{22}V_{31}|^2)$. The \CP\,violation is encoded in the imaginary part and in this case we define $J={\rm Im\,}(V_{22}V_{21}^*)^* (V_{32}V_{31}^*) $. If we take $(V_{22}V_{21}^*)$ and $(V_{32}V_{31}^*)$ as two sides of a triangle, they are elements of  $(V_{i2}V_{i1}^*)$ for $i=2$ and $3$. So, three sides of a Jarlskog triangle corresponds to three values for $i=1,2,3$, and these three complex numbers make up a triangle in the complex plane because $\sum_i V_{i2}V_{i1}^* =0$ due to  the unitarity of $V$. As we choose 2 and 1 from the column entries, there are three ways to make  column-triangles. Similarly, one can make three row-triangles which we do not use in this paper.
The physical magnitude of weak \CP\,violation in the CKM matrix  is given by the Jarlskog determinant $J^q$ which is twice of the area of the  Jarlskog triangle shown in Fig. \ref{fig:JTriPDG} where $\alpha,\beta,$ and $\gamma$ are the angles determined from hadron phenomena \cite{PDG18ckm}.

  But a simple form, readable from the $3\times 3$  CKM matrix itself, is given by \cite{KimSeo12}
  \begin{equation}
J=|{\rm Im}\,V_{31}V_{22}V_{13}|,~\textrm{after making Det.}V~{\rm real},\label{eq:JKSform}
 \end{equation}
 If the determinant is real as required, there is no imaginary part in
 \begin{equation}
 -\sum_{i,j,k}\epsilon_{ijk}\,({\rm Im}\,V_{i1}V_{j2}V_{k3})=0,\label{eq:KSform}
 \end{equation}
 \ie the permutations of $\{i,j,k\}=\{1,2,3\}$ add up the imaginary parts to zero, implying any set of $\{i,j,k\}$ has the same magnitude. So, any set out of 6 can be used as $J$, and there can be a consistency check on the determination of the CKM parameters by calculating $J$ from these 6 sets, \ie
\begin{equation}
 J=|\epsilon_{ijk}\,({\rm Im}\,V_{i1}V_{j2}V_{k3})|,{\rm with}~i,j,k\textrm{ not summed}.\label{eq:KSgeneral}
 \end{equation}
 As emphasized here, $J$ makes sense only if we use a unitary matrix $V$. It is useful to check the six terms independently so that the unitarity condition is satisfied. 
\begin{figure}[!t]
\hskip 0.01cm \includegraphics[width=0.35\textwidth]{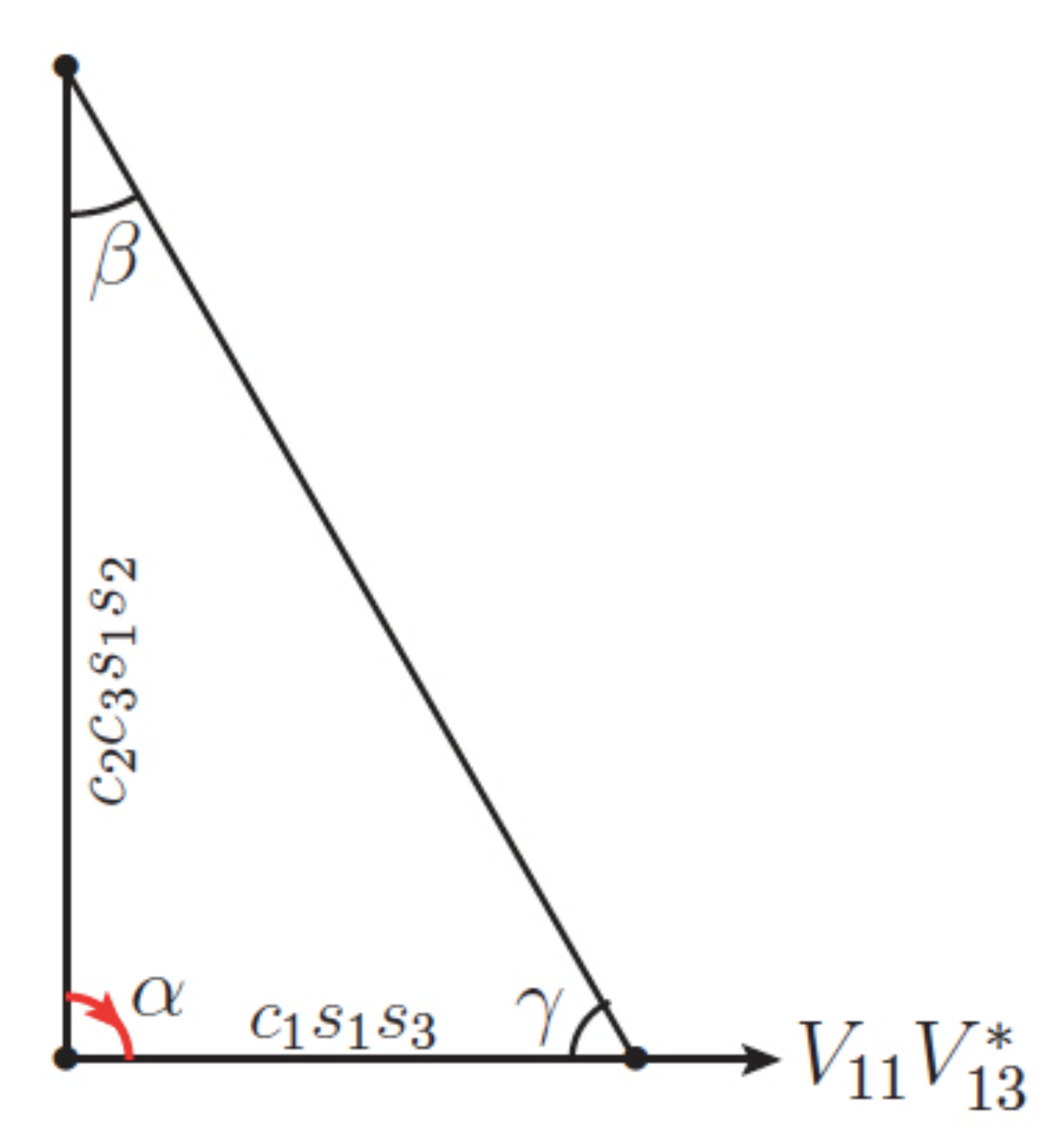}
\caption{The layout of the Jarlskog triangle in the KS form \cite{KimSeo11}. }\label{fig:JTriKS}
\end{figure}
 
\begin{figure}[!t]
\hskip 0.1cm \includegraphics[width=0.6\textwidth]{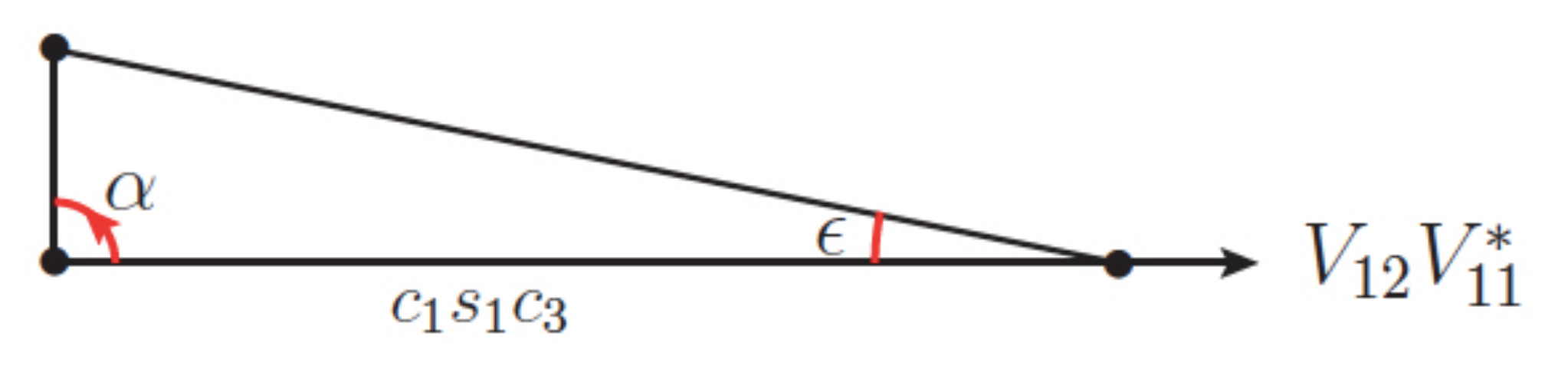}
\caption{The Jarlskog triangle with the horizontal axis $V^{\rm KS}_{12}V^{\rm KS\,*}_{11}$.}\label{fig:JTriabThin}
\end{figure}

 With the KS form, the Jarlskog triangle is shown in Fig. \ref{fig:JTriKS}. Note that the horizontal axis is the number  $V_{11}V_{13}^*$ which is real and hence it is sitting on the $x$-axis. Namely, in the KS form, one side of any Jarlskog triangle is sitting on the  $x$-axis. Out of three numbers from $V_{i1}V_{i3}^*$, if we take $i=3$ and using $V_{33}\simeq 1$, $V_{31}$ with the phase $\delks$ is the angle at the origin. This invariant angle appears in any Jarlskog triangle. If we consider the Jarlskog triangle  $V_{i2}V_{i1}^*$, we note $|V_{12}V_{11}^*|\simeq \lambda$ and  $|V_{22}V_{21}^*|\simeq \lambda$ and we have a shape shown in Fig. \ref{fig:JTriabThin}.
  Since $J=$O$(\lambda^{5})$, the small side has a length at most O($\lambda^4$), which implies that two angles are close to 90$^{\rm o}$.  Namely, from  trigonometry, if we have two long sides of length $\lambda$ and $\lambda+O(\lambda^4)$, the angle between them, $\epsilon$, is given for $\lambda\simeq 0.225$ by
 \begin{equation}
\cos\epsilon=1-O(\lambda^3)+\cdots \lesssim 0.9886\to \epsilon\lesssim 8.65^{\rm o}.
\end{equation}
One among $\alpha,\beta,$ and $\gamma$ must be $\delks$, and  $\epsilon$ cannot be  $\delks$ since there is no angle close to 0 among angles $\alpha,\beta,\gamma$ of Fig. \ref{fig:JTriPDG}.  The shape of this thin triangle is shown in Fig. \ref{fig:JTriabThin}, with an exaggerated $\epsilon$.
 
Our form of the CKM and PMNS matrices are  based on\footnote{The usual definition in Ceccucci  et al.  on the CKM matrix \cite{PDG18ckm} is the same as ours   but the definition on the PMNS matrix  in  Nakamura and Petcov  for the PMNS matrix \cite{PetcovPDG18} is the opposite to ours. } 
\begin{eqnarray}
&&V_{\rm CKM} =(U^{(u)}) (U^{(d)})^\dagger\label{eq:Vckm}\\ 
&&V_{\rm MNS} =(U^{(\nu)})  (U^{(e)})^\dagger\label{eq:Vmns}
\end{eqnarray}
where $U^{(u,d)}$ and $U^{(\nu,e)}$ are diagonalizing unitary matrices of L-handed  quark and lepton fields, \ie $u_L^{(m)}= U^{(u)}u_L^{(w)}$ and $d_L^{(m)}= U^{(d)}d_L^{(w)}$, respectively, for the following definition of mass terms defined on the weak eigenstates,
\begin{eqnarray}
{\cal H}=\bar{u}_L^{(w)} M_{u}^{(w)}u_R^{(w)}+\bar{d}_L^{(w)} M_{d}^{(w)}d_R^{(w)}+{\rm H.c.}
\end{eqnarray}
Gauge interactions for the charge raising operators give the CKM and PMNS matrices of (\ref{eq:Vckm}) and  (\ref{eq:Vmns}).

\section{Data on flavor physics}\label{sec:Data}
  
Our final results will be presented in the CKM and PMNS matrix forms such that  possible symmetries from these matrices  can be looked for.
 
\subsection{CKM matrix} \label{subsec:CKMdata} 

The CKM data in the PDG book was fitted to an approximate unitary matrix \cite{Wolfenstein83}, which is not adequate in calculating the Jarlskog determinant because an exact unitary matrix was not used. As we will see, $J^q$ is of order $10^{-5}$ and the approximate form of  \cite{Wolfenstein83} violates the unitarity condition at order $\lambda^4\simeq 2.6\times 10^{-3}$. Since the data does not satisfy the exact unitarity condition, we process the data such that the median values of the processed data satisfy the exact unitarity condition. Let us try to include as many data points as possible. [If the processed data generates too large error bars, then we will conclude that the BSM contribution is significant.] To remove the uncertainty on the phase, we choose a point if the absolute value of the  point does not have the phase dependence.

Let us express the median angles and the errors as
\begin{eqnarray}
&\theta_1=\bar{\theta}_1+\delta_1   ,~\theta_2=\bar{\theta}_2+\delta_2  ,~\bar{\theta}_3+\delta_3 ,~\alpha_{\rm KS}=\bar{\alpha}_{\rm KS}+\delta_{\rm KS},    \\
&\theta_{12}=\bar{\theta}_{12}+\delta_{12}  ,~\theta_{23}=\bar{\theta}_{23}+\delta_{23} ,~\bar{\theta}_{13}+\delta_{13} ,~\alpha_{\rm pdg}=\bar{\alpha}_{\rm pdg}+\delta_{\rm pdg}. 
\end{eqnarray}

We choose the following data entries for
\dis{
\textrm{KS form } :
(11)= &0.9742\pm 0.00021,~(12)=0.2243\pm 0.0005,~|(13)|=0.00394\pm 0.00036,\\
&(21)=-0.218\pm 0.004,~~|(31)|=0.0081\pm 0.0005 ,\label{eq:Input3} 
}
\dis{
\textrm{PDG form} :
(11)= &0.9742\pm 0.00021,~(12)=0.2243\pm 0.0005,~|(13)|=0.00394\pm 0.00036,\\
&(23)=0.0422\pm 0.008, ~|(33)|=1.019\pm 0.025,\label{eq:Input4} 
}
from the PDG data given below \cite{PDG18ckm}. 
  \dis{V_{\textrm{CKM}}^{\textrm{data}}=
  \left(
\begin{array}{ccc}
 0.9742 \pm 0.00021 & 0.2243 \pm 0.0005 & 0.00394 \pm 0.00036 \\
 -0.218 \pm 0.004 & 0.997 \pm 0.017 & 0.0422 \pm 0.008 \\
 0.0081 \pm 0.0005 & 0.0394 \pm 0.0023 & 1.019 \pm 0.025 \\
\end{array}
\right).
}

After processing, we got the angles, firstly using row entries 
 \begin{eqnarray}
& \bar{\theta}_1^{\rm KS , Row}=12.7853 ^{\rm o} ,~ \bar{\theta}^{\rm KS , Row}_2=2.1279^{\rm o} ,~
 \bar{\theta}_3^{\rm KS , Row}=1.00634^{\rm o} ,\label{eq:rowKS}\\
&\delta_1^{\rm KS , Row}= \pm 0.621641^{\rm o},~ 
\delta_2^{\rm KS, Row}= \pm 0.136906^{\rm o},~
\delta_3^{\rm KS, Row}=  \pm 0.130049^{\rm o},\label{eq:rowKS}
\end{eqnarray}
and second using column entries
 \begin{eqnarray}
& \bar{\theta}_1^{\rm KS , 1Column}=12.7921 ^{\rm o} ,~ \bar{\theta}^{\rm KS , 1Column}_2=2.1279^{\rm o} ,~
 \bar{\theta}_3^{\rm KS , 1Column}=1.00634^{\rm o} ,\label{eq:1columnKS}\\
&\delta_1^{\rm KS , 1Column}= \pm 0.407459^{\rm o},~ 
\delta_2^{\rm KS , 1Column}= \pm 0.193615^{\rm o},~
\delta_3^{\rm KS , 1Column}=  \pm 0.0919583^{\rm o}.\label{eq:1columnKS}
\end{eqnarray}
Averaging these, we obtained the following KS angles and $J_{\rm KS}^q$ ,
 \begin{eqnarray}
& \bar{\theta}_1^{\rm KS}=12.79 ^{\rm o} ,~ \bar{\theta}^{\rm KS}_2=2.1279^{\rm o} ,~
 \bar{\theta}_3^{\rm KS}=1.00634^{\rm o} ,\label{eq:averageKS}\\
&\delta_1^{\rm KS}= \pm 0.340779^{\rm o},~ 
\delta_2^{\rm KS}= \pm 0.111783^{\rm o},~
\delta_3^{\rm KS}=  \pm 0.0750837^{\rm o},\label{eq:averageerrorsKS}\\
&J_{\rm KS}^q=(3.11403\pm0.325)\times10^{-5}|\sin\delks|.\label{eq:aJKS}
\end{eqnarray}
Note that the error propagation during the procedure adds extra systematic errors in addition to the errors in PDG data.
Then, the evaluated KS form of the CKM matrix determined from data is
\dis{
V_{\rm CKM}^{\rm KS}=
\begin{pmatrix}
 0.975188\pm 0.0013167 \,, & 0.221345\pm 0.00579925\,, & 0.003888\pm 0.000307\\[1.5em]
-0.221226\pm 0.00579616\,,  & \begin{array}{l}0.974365\pm 0.00131633 \\[0.2em] 
 -i(0.00065\pm 0.000003)\sin\delks\\[0.2em]
  +(0.00065\pm 0.000003)\cos\delks \end{array}, &  \begin{array}{l} +0.01712  \pm 0.000002\\[0.2em] 
  +i(0.03712\pm 0.001473)\sin\delks\\[0.2em]
  -(0.03712\pm 0.001473)\cos\delks \end{array}   \\[2.2em]
  \begin{array}{c}   -i(0.00822\pm 0.000482)\sin\delks \\
  -(0.00822\pm 0.000482)\cos\delks
   \end{array} , & \begin{array}{l} -0.017551\pm 0.000002\\[0.2em] +i(0.03620\pm 0.001379)\sin\delks\\[0.2em] +(0.03620\pm 0.001379)\cos\delks \end{array}, & \begin{array}{l} +0.999156\pm 0.00007\\[0.2em] + i(0.00064\pm 0.00005)\sin\delks \\[0.2em] + (0.00064\pm 0.00005)\cos\delks \end{array} 
\end{pmatrix} .\label{eq:aCKMdata}
}

In the same way, we obtain the following from Eq. (\ref{eq:Input4}) and  the PDG parametrization \cite{PDG18},
\begin{eqnarray}
& \bar{\theta}_{12}^{\rm  PDG, Row}=12.9658^{\rm o} ,~ \bar{\theta}_{23}^{\rm  PDG, Row}=2.37144^{\rm o} ,~
 \bar{\theta}_{13}^{\rm  PDG, Row}= 0.225815^{\rm o},\\
 & \delta_{12}^{\rm  PDG, Row}= \pm1.61928^{\rm o},~ 
\delta_{23}^{\rm  PDG, Row}= \pm 0.452794^{\rm o},~
\delta_{13}^{\rm  PDG, Row}=  \pm 0.0206391 ^{\rm o},\label{eq:rowCKMangles}
\end{eqnarray} 
from the row entries, and
\begin{eqnarray}
& \bar{\theta}_{12}^{\rm  PDG, 3Column}=12.9658^{\rm o} ,~ \bar{\theta}_{23}^{\rm  PDG}=2.37144^{\rm o} ,~
 \bar{\theta}_{13}^{\rm  PDG, 3Column}= 0.221345^{\rm o},\\
 & \delta_{12}^{\rm  PDG, 3Column}= \pm0.0280565^{\rm o},~ 
\delta_{23}^{\rm  PDG, 3Column}= \pm 0.723851^{\rm o},~
\delta_{13}^{\rm  PDG, 3Column}=  \pm 0.047828 ^{\rm o},\label{eq:3columnCKMangles}
\end{eqnarray} 
from the third column entries. Note that the third column has large error bars for the (23) and (33) elements  regardless of the extra systematic error from the process. By  averaging these, the following PDG angles and $J_{\rm PDG}^q$ are obtained,
\begin{eqnarray}
& \bar{\theta}_{12}^{\rm  PDG}=12.9658^{\rm o} ,~ \bar{\theta}_{23}^{\rm  PDG}=2.37144^{\rm o} ,~
 \bar{\theta}_{13}^{\rm  PDG}= 0.225113^{\rm o},\\
 & \delta_{12}^{\rm  PDG}= \pm0.0280523^{\rm o},~ 
\delta_{23}^{\rm  PDG}= \pm 0.383876^{\rm o},~
\delta_{13}^{\rm  PDG}=  \pm 0.01895 ^{\rm o},\label{eq:averageCKMangles}\\
&J_{\rm PDG}^q=(3.5515\pm0.647)\times10^{-5}|\sin\alpha_{_{\rm  PDG}}|. \label{eq:JPDG}
\end{eqnarray} 
Then, the evaluated KS form of the CKM matrix determined from data is
\dis{
V_{\rm CKM}^{\rm PDG}=
\left(
\begin{array}{ccc}
 0.974497\pm 0.000110 & 0.224368\pm 0.000477& 
\begin{array}{c}
 - i (0.003929\pm 0.000331)\sin{\delpdg} \\
+(0.003929\pm 0.000331)\cos{\delpdg} \\
\end{array}
\\[0.5em]
\begin{array}{c}
 -0.224178\pm 0.000480 \\
- i(0.000158\pm 0.000003)\sin{\delpdg}  \\
 -  (0.000158\pm 0.000003)\cos{\delpdg} \\
\end{array}
& 
\begin{array}{c}
 0.973669\pm 0.000292 \\
- i (0.000036\pm 0.000005)\sin{\delpdg} \\
 - (0.000036\pm 0.000005)\cos{\delpdg} \\
\end{array}
 & 
\begin{array}{c}
 0.041377\pm 0.006694 \\
\end{array}
\\[0.5cm]
\begin{array}{c}
 0.009284\pm 0.001467\ \\
 -i(0.0038255\pm 0.000001) \sin{\delpdg}  \\
 - (0.0038255\pm 0.000001)\cos{\delpdg} \\
\end{array}
 &
\begin{array}{c}
 -0.040323\pm 0.006523 \\
 -i(0.000881\pm 0.0000002) \sin{\delpdg} \\
 - (0.000881\pm 0.0000002) \cos{\delpdg} \\
\end{array}
 &
\begin{array}{c}
 0.999136\pm 0.000277
\end{array}
\\
\end{array}
\right) .\label{eq:aPDGdata}
}

From the invariance of Jarlskog determinant,
\dis{
\left|\frac{\sin\delpdg}{\sin\delks}\right|=\frac{(3.11403\pm0.325)}{(3.5515\pm0.647)}=0.88\pm 0.18,\label{eq:Ratiopdgks}
}
leading to 
\dis{
\alpha_{\rm PDG}=(61\pm22)^{\rm o}\label{eq:Anglepdg}
}
for $ \alpha_{\rm KS}=90^{\rm o}$,
where the large error bar is primarily due to the large error of $J_{\rm PDG}^q$ in Eq. (\ref{eq:JPDG}). 

  Comparing $(61\pm22)^{\rm o}$ with the value that the PDG book determined $\gamma=\left(73.5^{+4.2}_{-5.1}\right)^{\rm o} $,  and   Eq. (\ref{eq:aJKS}) with $J^q_{\rm PDG}=(3.18\pm 0.15)\times 10^{-5}$, we conclude that they are consistent. But, we emphasize that our method follows the unitary matrix in the whole analysis and should be followed in the future which is contrasted to the method  in the PDG book using   the approximate Wolfenstein parametrization \cite{Wolfenstein83}.

\subsection{Mass matrix of quarks} \label{subsec:QuarkMass} 

Let us diagonalize $\Qem=-\frac13$ quarks. Then, the CKM matrix is $U^{(u)}$. The mass matrix for  $\Qem=+\frac23$ quarks is
\begin{eqnarray}
{\cal H}=\bar{u}_R^{(m)} M_{u}^{(\rm diag.)}u_L^{(m)}+{\rm H.c.}=\bar{u}_R^{(w)} V^{(u)\dagger} M_{u}^{(\rm diag.)} U^{(u)}u_L^{(w)}+{\rm H.c.}=\bar{u}_R^{(w)} V^{(u)\dagger} M_{u}^{(\rm diag.)} V_{\rm CKM}u_L^{(w)}+{\rm H.c.}
\end{eqnarray}
Therefore, the mass matrix one can write the following, supported by the data, as
\begin{equation}
H_{\rm mass}^{(w)}
  =m_t ~V^{(u)\dagger} \begin{pmatrix} \frac{m_u}{m_t}&0&0\\ 0&\frac{m_c}{m_t}&0 \\ 0&0& 1\end{pmatrix} V_{\rm CKM}^{\rm data}.\label{eq:DefMass}
\end{equation}
Depending on the identification of the right-handed quark singlets matrix $u_R$, we obtain the hierarchical terms in the mass matrix $H_{\rm mass}^{(w)}$ in Eq. (\ref{eq:DefMass}) for which we may use  some idea on discrete symmetries.

\color{black}

\subsection{Neutrino oscillation and MNS matrix} \label{subsec:MNSdata} 
   
 The PDG book gives the values in the PDG parametrization and we can  convert $V^{\rm PDG}_{\rm PMNS}$ to the values in the KS form $ V^{\rm KS}_{\rm PMNS}$ in the following way,   
\dis{ 
V^{\rm PDG}_{\rm PMNS}\to V^{\rm KS}_{\rm MNS}=L\,
 V^{\rm PDG}_{\rm PMNS} \begin{pmatrix} 1&0&0\\ 0&1&0 \\ 0&0& e^{i\delta^l_{\rm PDG}}
\end{pmatrix},
} 
where 
\dis{
L=\left(
\begin{array}{ccc}
 1 & 0 & 0 \\
 0 & \frac{(C_{23} S_{12}+C_{12}S_{13} S_{23} e^{-i \delta_{\rm PDG}^l })}{\sqrt{2 C_{12} C_{23}S_{13} S_{23} S_{12} \cos \delta_{\rm PDG}^l +C_{23}^2 S_{12}^2+C_{12}^2 S_{13}^2 S_{23}^2}} & 0 \\
 0 & 0 & \frac{\sqrt{2 C_{12} C_{23} S_{13} S_{23} S_{12} \cos \delta_{\rm PDG}^l +C_{23}^2 S_{12}^2+C_{12}^2 S_{13}^2 S_{23}^2}}{C_{12} S_{13} S_{23}+C_{23} S_{12} e^{i \delta_{\rm PDG}^l }} \\
\end{array}
\right).
}
Then, the real entries in the KS form are 
\dis{
&V^{\rm KS}_{\rm PMNS(11)}=  C_{12} C_{13} , \\
&V^{\rm KS}_{\rm PMNS(12)}= C_{13} S_{12}, \\
&V^{\rm KS}_{\rm PMNS(13)}= S_{13},
\\
&V^{\rm KS}_{\rm PMNS(21)}=\textcolor{red}{-}\sqrt{C_{12} C_{23} S_{13} S_{23} S_{12} e^{-i \delta ^l{}_{\text{PDG}}} \left(1+e^{2 i \delta ^l{}_{\text{PDG}}}\right)+C_{23}^2 S_{12}^2+C_{12}^2 S_{13}^2 S_{23}^2},\\
&V^{\rm KS}_{\rm PMNS(31)}=\frac{\left(S_{12} S_{23}\textcolor{red}{-}C_{12} C_{23} S_{13} e^{i \delta ^l{}_{\text{PDG}}}\right) \sqrt{ C_{12} C_{23} S_{13} S_{23} S_{12} \cos{ \delta ^l{}_{\text{PDG}}}+C_{23}^2 S_{12}^2+C_{12}^2 S_{13}^2 S_{23}^2}}{C_{12} S_{13} S_{23}+C_{23} S_{12} e^{i \delta ^l{}_{\text{PDG}}}},
} 
where $\Theta_{ij}$ are the real angles in the PDG parametrization. 
 
Now, as in the CKM case let us use the \CP--phase independent absolute values of  data, the  (11), (12), (13), (21), and (31) elements in the KS form.\footnote{Since there is an overall $\mathcal{CP}$ phase factor, we can work out with the absolute value of the (31) element.} Since atmospheric neutrino data generically are difficult to analyze outside the experimental collaborations, the 2019 data on   $V_{\rm PMNS}$ from NuFIT  \cite{NuFIT18} gives the matrix elements\footnote{Since the data are given in terms of the angles, the unitarity conditions are automatically satisfied.} excluding and including the Super-Kamiokande atmospheric (SK-atm) data.\footnote{A theoretical fit with a tribimaximal mixing is given in \cite{Valle18}.}  Therefore, we present for these two cases separately.

\subsubsection{\bf Excluding the SK-atm data}
\centerline{\underline{\it Normal hierarchy}}
 \dis{
(11)_{\textrm{NH}}^{\textrm{No SK-atm}}=&0.821427^{+0.007497}_{-0.007305},~ (12)_{\textrm{NH}}^{\textrm{No SK-atm}}=0.550313^{+0.011184}_{-0.010898},~   (13)_{\textrm{NH}}^{\textrm{No SK-atm}}=0.149708^{+0.002243}_{-0.002243},~ \\ &(21)_{\textrm{NH}}^{\textrm{No SK-atm}}=0.288311^{+0.048803}_{-0.036610},~|(31)|_{\textrm{NH}}^{\textrm{No SK-atm}}=0.492071^{+0.374396}_{-0.271582}.
}

The leptonic angles without SK-atm contribution and with normal ordered mass are obtained as 
 \begin{eqnarray}
&&\theta_1^{\rm KS}=(\ar{34.7721}{0.753102 }{0.733896}) ^{\rm o}
 ,~ \theta^{\rm KS}_2=(\ar{59.4466}{1.07601}{1.22664}) ^{\rm o},~
 \theta_3^{\rm KS}=(\ar{15.2185}{1.55645}{1.51545}) ^{\rm o}
\end{eqnarray}
In this case, we obtain  from any one out of 6 possible products of the form given in Eq. (\ref{eq:KSgeneral}),
\begin{equation}
J=(\ar{2.96}{0.29}{0.29})\times10^{-2}|\sin{\alpha _{\text{KS}}}|.\label{JXSKN}
\end{equation}
The PMNS matrix becomes
\begin{equation}
\left(
\begin{array}{ccc}
\ar{0.821427}{0.00749625}{0.00730508}, & \ar{0.550313}{0.0111839}{0.0108975}, &  \ar{0.149708}{0.0152156}{0.0148153} ,
 \\[0.5cm]
 - \ar{0.289913}{0.0107329}{0.0117967} ,
 & \begin{array}{c} \ar{0.402922}{0.00525912}{0.0061485}\\[0.5em] -i( \ar{0.226055}{0.0188886}{0.0171532}) \sin{\alpha _{\text{KS}}}\\[0.5em]+( \ar{0.226055}{0.0188886}{0.0171532}) \cos{\alpha _{\text{KS}}} \end{array},
 & \begin{array}{c} \ar{0.109611}{0.0102309}{0.00989485}\\[0.5em]+i( \ar{0.830957}{0.00970954}{0.0106587}) \sin{\alpha _{\text{KS}}}\\[0.5em]-( \ar{0.830957}{0.00970954}{0.0106587})\cos{\alpha _{\text{KS}}} \end{array},
 \\[1cm]
 \begin{array}{c}-( \ar{0.491128}{0.0107746}{0.0109827})\cos{\alpha _{\text{KS}}}\\[0.5em]-i(\ar{0.491128}{0.0107746}{0.0109827}) \sin{\alpha _{\text{KS}}} \end{array},
 & \begin{array}{c}- \ar{0.13344}{0.0119605}{0.0117196}\\[0.5em] +i( \ar{0.682571}{0.00830546}{0.00899913}) \sin{\alpha _{\text{KS}}}\\[0.5em]+(\ar{0.682571}{0.00830546}{0.00899913})\cos{\alpha _{\text{KS}}}  \end{array},
 & \begin{array}{c}+ \ar{0.490514}{0.00792606}{0.0101431}\\[0.5em] +i( \ar{0.185687}{0.012533}{0.0103979}) \sin{\alpha _{\text{KS}}}\\[0.5em]+( \ar{0.185687}{0.012533}{0.0103979})\cos{\alpha _{\text{KS}}} \end{array},
  \\
\end{array}
\right)\label{eq:XSKN}
\end{equation} 
 
\bigskip
\centerline{\underline{\it Inverted hierarchy}}
 \dis{
(11)_{\textrm{IH}}^{\textrm{No SK-atm}}=&0.82134^{+0.007496}_{-0.007304},~ (12)_{\textrm{IH}}^{\textrm{No SK-atm}}=0.550255^{+0.011183}_{-0.010896},~   (13)_{\textrm{IH}}^{\textrm{No SK-atm}}=0.150398^{+0.002243}_{-0.002243},~ \\ &(21)_{\textrm{IH}}^{\textrm{No SK-atm}}=0.393394^{+0.040942}_{-0.043898},~|(31)|_{\textrm{IH}}^{\textrm{No SK-atm}}=0.413087^{+0.179299}_{-0.192553}, 
}
The leptonic angles without SK-atm contribution and with inversely ordered mass are obtained as
 \begin{eqnarray}
&&\bar{\theta}_1^{\rm KS}=(\ar{34.7808}{0.75294 }{0.733662}) ^{\rm o}
 ,~\bar{\theta}^{\rm KS}_2=(\ar{46.3988}{0.794642}{0.903062}) ^{\rm o},~
\bar{\theta}_3^{\rm KS}=(\ar{15.287}{1.55488}{1.51472}) ^{\rm o}
\end{eqnarray}
In this case, we obtain  from any one out of 6 possible products of the form given in Eq. (\ref{eq:KSgeneral}),
\begin{equation}
J=(\ar{3.39}{0.33}{0.32})\times10^{-2}|\sin{\alpha _{\text{KS}}}|.\label{JXSKI}
\end{equation}
The MNS matrix becomes
\begin{equation}
\left(
\begin{array}{ccc}
 \ar{0.82134}{0.00749629}{0.00730436}, &  \ar{0.550255}{0.011183}{0.0108963}, &\ar{0.150398}{0.0152014}{0.014809},
\\[0.5cm]
 - \ar{0.393394}{0.00939307}{0.00974662}, &\begin{array}{c}\ar{0.546384}{0.00565927}{0.00606947 }
 \\[0.5em]-i(\ar{0.190927}{0.0171211}{0.0162823 }) \sin{\alpha _{\text{KS}}}
 \\[0.5em] +(\ar{0.190927}{0.0171211}{0.0162823 })\cos{\alpha _{\text{KS}}}
 \end{array},   &\begin{array}{c}\ar{0.14934}{0.0130298}{0.0121399 }\\ [0.5em]
+ i(\ar{0.698534}{0.00743625}{0.00814051 }) \sin{\alpha _{\text{KS}}}
 \\[0.5em]-(\ar{0.698534}{0.00743625}{0.00814051 }) \cos{\alpha _{\text{KS}}} \end{array},
\\[1cm]
\begin{array}{c}   
 -(\ar{0.413087 }{0.00953209}{0.00982088 })\cos{\alpha _{\text{KS}}}
 \\[0.5em]-i(\ar{0.413087 }{0.00953209}{0.00982088 }) \sin{\alpha _{\text{KS}}} 
 \end{array}& \begin{array}{c}
 -\ar{0.181825}{0.0162895 }{0.0155377}
 \\[0.5em]
+i(\ar{0.573734}{0.00594818}{0.00636034}) \sin{\alpha _{\text{KS}}}
 \\[0.5em]+(\ar{0.573734}{0.00594818}{0.00636034}) \cos{\alpha _{\text{KS}}}
 \end{array} & \begin{array}{c}\ar{0.665234}{0.00718007}{0.00791545}\\[0.5em]+ i(\ar{0.156816}{0.0134947}{0.0124851}) \sin{\alpha _{\text{KS}}}\\[0.5em]+(\ar{0.156816}{0.0134947}{0.0124851})\cos{\alpha _{\text{KS}}} \end{array}\\
\end{array}
\right)\label{eq:XSKI}
\end{equation} 

\bigskip
\subsubsection{\bf Including  SK-atm data}
\centerline{\underline{\it Normal hierarchy}}
 \begin{eqnarray}
&(11)_{\textrm{NH}}^{\textrm{SK-atm}}=0.821427^{+0.007496}_{-0.007305},~ (12)_{\textrm{NH}}^{\textrm{SK-atm}}=0.550313^{+0.011183}_{-0.010898},~   (13)_{\textrm{NH}}^{\textrm{SK-atm}}=0.149708^{+0.002071}_{-0.002243},~ \\ &(21)_{\textrm{NH}}^{\textrm{SK-atm}}=0.289913^{+0.050638}_{-0.036595},~|(31)|_{\textrm{NH}}^{\textrm{SK-atm}}=0.491128^{+0.370763}_{-0.259686}, 
\end{eqnarray}
The leptonic angles with SK-atm contribution and normal ordered mass are obtained as 
  \begin{eqnarray}
&&\bar{\theta}_1^{\rm KS}=(\ar{34.7721}{0.753102 }{0.733896}) ^{\rm o}
 ,~ \bar{\theta}^{\rm KS}_2=(\ar{59.4466}{1.07601}{1.22664}) ^{\rm o},~
\bar{\theta}_3^{\rm KS}=(\ar{15.2185}{1.55645}{1.51545}) ^{\rm o}
\end{eqnarray}
In this case, we obtain  from any one out of 6 possible products of the form given in Eq. (\ref{eq:KSgeneral}),
\begin{equation}
J=(\ar{2.96}{0.29}{0.29})\times10^{-2}|\sin{\alpha _{\text{KS}}}|
\end{equation}
The MNS matrix becomes
\begin{equation}
\left(
\begin{array}{ccc}
\ar{0.821427}{0.00749625}{0.00730508}, & \ar{0.550313}{0.0111839}{0.0108975}, &  \ar{0.149708}{0.0152156}{0.0148153} ,
 \\[0.5cm]
 - \ar{0.289913}{0.0107329}{0.0117967} ,
 & \begin{array}{c} \ar{0.402922}{0.00525912}{0.0061485}\\[0.5em] -i( \ar{0.226055}{0.0188886}{0.0171532}) \sin{\alpha _{\text{KS}}}\\[0.5em]+( \ar{0.226055}{0.0188886}{0.0171532}) \cos{\alpha _{\text{KS}}} \end{array},
 & \begin{array}{c} \ar{0.109611}{0.0102309}{0.00989485}\\[0.5em]+i( \ar{0.830957}{0.00970954}{0.0106587}) \sin{\alpha _{\text{KS}}}\\[0.5em]-( \ar{0.830957}{0.00970954}{0.0106587})\cos{\alpha _{\text{KS}}} \end{array},
 \\[1cm]
 \begin{array}{c}-( \ar{0.491128}{0.0107746}{0.0109827})\cos{\alpha _{\text{KS}}}\\[0.5em]-i(\ar{0.491128}{0.0107746}{0.0109827}) \sin{\alpha _{\text{KS}}} \end{array},
 & \begin{array}{c}- \ar{0.13344}{0.0119605}{0.0117196}\\[0.5em] +i( \ar{0.682571}{0.00830546}{0.00899913}) \sin{\alpha _{\text{KS}}}\\[0.5em]+(\ar{0.682571}{0.00830546}{0.00899913})\cos{\alpha _{\text{KS}}}  \end{array},
 & \begin{array}{c}+ \ar{0.490514}{0.00792606}{0.0101431}\\[0.5em] +i( \ar{0.185687}{0.012533}{0.0103979}) \sin{\alpha _{\text{KS}}}\\[0.5em]+( \ar{0.185687}{0.012533}{0.0103979})\cos{\alpha _{\text{KS}}} \end{array},
  \\
\end{array}
\right)\label{eq:OSKN}
\end{equation}

\bigskip
\centerline{\underline{\it Inverted hierarchy}}
 \begin{eqnarray}
&(11)_{\textrm{IH}}^{\textrm{SK-atm}}=0.82134^{+0.007496}_{-0.007208},~ (12)_{\textrm{IH}}^{\textrm{SK-atm}}=0.550255^{+0.011183}_{-0.010753},~   (13)_{\textrm{IH}}^{\textrm{SK-atm}}=0.150398^{+0.002071}_{-0.002243},~ \\ &(21)_{\textrm{IH}}^{\textrm{SK-atm}}=0.38806^{+0.039041}_{-0.04355},~|(31)|_{\textrm{IH}}^{\textrm{SK-atm}}=0.418102^{+0.167767}_{-0.187833}, 
\end{eqnarray}
The leptonic angles with SK-atm contribution and inversely ordered mass are obtained as 
 \begin{eqnarray}
&&\bar{\theta}_1^{\rm KS}=(\ar{34.7808}{0.752861}{0.724023}) ^{\rm o}
 ,~ \bar{\theta}^{\rm KS}_2=(\ar{47.1341}{0.71626}{0.825659}) ^{\rm o},~
 \bar{\theta}_3^{\rm KS}=(\ar{15.287}{1.5557}{1.49464}) ^{\rm o}
\end{eqnarray}
In this case, we obtain  from any one out of 6 possible products of the form given in Eq. (\ref{eq:KSgeneral}),
\begin{equation}
J=(\ar{3.39}{0.33}{0.31})\times10^{-2}|\sin{\alpha _{\text{KS}}}|.
\end{equation}
The MNS matrix becomes
\begin{equation}
\left(
\begin{array}{ccc}
 \ar{0.82134}{0.0074955}{0.0072084}, &\ar{0.550255}{0.0111827}{0.010753}, & \ar{0.150398}{0.0152091}{0.0146127},
  \\[0.5cm]
 -\ar{0.38806}{0.0090123}{0.0092819} ,
 & \begin{array}{c}\ar{0.538975}{0.00529664}{0.00561701}\\[0.5em] -i(\ar{0.193244}{0.0175702}{0.0164689}) \sin{\alpha _{\text{KS}}}\\[0.5em]+(\ar{0.193244}{0.0175702}{0.0164689})\cos{\alpha _{\text{KS}}} \end{array},
 & \begin{array}{c}\ar{0.147315}{0.0132161}{0.0121528}\\[0.5em] i(\ar{0.707014}{0.00704164}{0.00761188}) \sin{\alpha _{\text{KS}}}\\[0.5em]-(\ar{0.707014}{0.00704164}{0.00761188})  \cos{\alpha _{\text{KS}}}\end{array},
  \\[1cm]
 \begin{array}{c} -(\ar{0.418102}{0.0092793}{0.0094415}) \cos{\alpha _{\text{KS}}}\\[0.5em]-i(\ar{0.418102}{0.0092793}{0.0094415}) \sin{\alpha _{\text{KS}}}\end{array},
  & \begin{array}{c}-\ar{0.17936}{0.0162496}{0.0152948}\\ [0.5em]+i(\ar{0.5807}{0.00572711}{0.0060482}) \sin{\alpha _{\text{KS}}}\\[0.5em]+(\ar{0.5807}{0.00572711}{0.0060482}) \cos{\alpha _{\text{KS}}} \end{array},
  & \begin{array}{c}\ar{0.656214}{0.00664119}{0.00725026}\\[0.5em]+i(\ar{0.158719}{0.014013}{0.012759}) \sin{\alpha _{\text{KS}}}\\[0.5em]+(\ar{0.158719}{0.014013}{0.012759})  \cos{\alpha _{\text{KS}}}\end{array},\\
\end{array}
\right)\label{eq:OSKI}
\end{equation}\color{black}
 
   
\bigskip
 
Equations (\ref{eq:XSKN}) and (\ref{JXSKN}) for the case without using the SK atmospheric data, which may be approximated  to a tribimaximal mixing. Our error bounds of the resulting $J^\ell$, Eq.  (\ref{JXSKN}), is roughly at a 1$\sigma$ from a model calculation of Ref. \cite{Ramond18}: their values are $J^\ell\sim 3.3\times 10^{-2}|\sin\delpdg|$ and the leptonic phase $\delpdg\sim \pm   238^{\rm o}$. This kind of comparison can be performed for Eqs. (\ref{eq:XSKI}, \ref{eq:OSKN}, \ref{eq:OSKI}) also. 
\subsection{Comparison of the CKM and PMNS phases}

Let us discuss with the KS parametrizations of the CKM and MNS matrices and their determination given in Eqs. (\ref{eq:aCKMdata}, \ref{eq:aJKS}) and  (\ref{eq:XSKN}, \ref{JXSKN}). We noted that $\delks\simeq 90^{\rm o}$. Reference \cite{NuFIT18} gives $J^\ell_{\rm best}=-0.019=3.33\times 10^{-2}\sin\delpdg^\ell$, leading to $\delpdg^\ell\simeq - 34.79^{\rm o}$.  For NH without KS, $\delpdg$ is given as 215$^{\rm o}$. The equivalence of $J^\ell$ in two parametrizations give, using the NH without the SK data, Eq. (\ref{JXSKN}),  
 and the central values
\dis{ 
(\ar{2.96}{0.29}{0.29})\times 10^{-2}\sin\delks=-(1.92_{-0.036}^{+0.037}) \times 10^{-2}   \to \delks^\ell\simeq -(\ar{40.4697}{4.94848}{4.86999})^{\rm o}.
} 
Obviously, $\delks\ne \pm\delks^\ell$.  
 
As in  Eq. (\ref{eq:DefMass}) in the quark sector, we may use  some idea on discrete symmetries on the mass matrix of neutrinos but here the discussion is more involved because the neutrino masses arise from dimension 5 operators. 
    
\section{Conclusion}\label{sec:Conclusion}
 We attempted to present the approximate CKM and PMNS matrices in the form of $3\times 3$ matrices,  Eqs. (\ref{eq:aCKMdata}) and (\ref{eq:XSKN},\ref{eq:XSKI},\ref{eq:OSKN},\ref{eq:OSKI}), by determining three real angles with 5 data inputs. In the final matrices, we  included the least known phases $\delks$ and $\delksl$ as free parameters.  The Jarlskog invariants in the quark and lepton sectors are determined as $J^q \sim 3.11\times 10^{-5}|\sin\delks|$ and $J^\ell\sim 2.96\times 10^{-2}|\sin\delksl|$ for NH, respectively, which are the essential information for   future BSM model buildings from string compactification.

\acknowledgments{J.E.K. thanks S. K. Kang, S. Khalil, J. W. F. Valle, and U. Yang for the helpful discussions.  This work is supported in part  by the National Research Foundation (NRF) grant  NRF-2018R1A2A3074631, and in addition M.S. is supported in part by the
Hyundai Motor Chung Mong-Koo Foundation.}
 

\end{document}